\documentclass[conference]{IEEEtran}
\usepackage[T1]{fontenc}
\usepackage{verbatim}
\usepackage{amsmath}
\usepackage{graphicx}
\usepackage{xcolor}
\usepackage[colorlinks=true, linkcolor=darkcyan, citecolor=black, urlcolor=darkcyan, pdfborder={0 0 0}]{hyperref}    

\makeatletter
\usepackage{amssymb}          

\def\BibTeX{{\rm B\kern-.05em{\sc i\kern-.025em b}\kern-.08em
    T\kern-.1667em\lower.7ex\hbox{E}\kern-.125emX}}
\usepackage{balance}
\usepackage{cite}
\makeatother

\definecolor{darkcyan}{rgb}{0.0, 0.35, 0.55}

\definecolor{darkgreen}{rgb}{0.0, 0.5, 0.0}

\newcommand{\yspace}{-0 pt}

\usepackage{fancyhdr}

\fancypagestyle{firstpage}{%
  \fancyhf{} 
  
  \fancyhead[L]{\scriptsize \copyright~ 2025 IEEE. Personal use of this material is permitted.  Permission from IEEE must be obtained for all other uses, in any current or future media, including reprinting/republishing this material for advertising or promotional purposes, creating new collective works, for resale or redistribution to servers or lists, or reuse of any copyrighted component of this work in other works. 
  DOI: 10.1109/DSP65409.2025.11074995}%
}

\begin{document}
\title{Joint Sampling Frequency Offset Estimation and Compensation Based on the Farrow Structure}

\author{\IEEEauthorblockN{Deijany Rodriguez Linares}
	\IEEEauthorblockA{Link\"oping University\\
		Department of Electrical Engineering\\
		581  83 Link\"oping, Sweden\\
		Email: deijany.rodriguez.linares@liu.se}
	\and
	\IEEEauthorblockN{Oksana Moryakova}
	\IEEEauthorblockA{Link\"oping University\\
		Department of Electrical Engineering\\
		581  83 Link\"oping, Sweden\\
		Email: oksana.moryakova@liu.se}
		\and
	\IEEEauthorblockN{H\aa kan Johansson}
	\IEEEauthorblockA{Link\"oping University\\
		Department of Electrical Engineering\\
		581  83 Link\"oping, Sweden\\
		Email: hakan.johansson@liu.se}
}
\maketitle
\thispagestyle{firstpage} 

\begin{abstract}
	This paper introduces a sampling frequency offset (SFO) estimation method based on the Farrow structure, which is typically utilized for the SFO compensation and thereby enables a reduction of the implementation complexity of the SFO estimation. The proposed method is implemented in the time domain and works for arbitrary bandlimited signals, thus with no additional constraints on the waveform structure. Moreover, it can operate on only the real or imaginary part of a complex signal, which further reduces the estimation complexity. Furthermore, the proposed method can simultaneously estimate the SFO and additional sampling time offset (STO) and it is insensitive to other synchronization errors, like carrier frequency offset. Both the derivations of the proposed method and its implementation are presented, and through simulation examples, it is demonstrated that it can accurately
	estimate both SFO and STO for different types of bandlimited signals.
\end{abstract}
\vspace{\yspace}
\section{Introduction\label{sec:Intro}}
In digital communication systems, accurate synchronization is crucial for the correct reception and interpretation of signals. Among various synchronization challenges, sampling frequency offset (SFO) critically impacts the system performance as it leads to inter-carrier and inter-symbol interference \cite{Matin_2021, Wang_2003}. Most of the existing methods for SFO estimation are designed for orthogonal frequency division multiplexing (OFDM) signals, and therefore carry out the estimation in the frequency domain \cite{Hou_2020,Li_2024, Nguyen_2009, Tsai_2005, Kim_2011, Xu2014, WU2024}. Several methods consider joint estimation of SFO and carrier frequency offset (CFO) due to their interrelated effects \cite{Nguyen_2009, Tsai_2005, Kim_2011, Xu2014, WU2024}. Alternatively, time-domain SFO estimation methods have been proposed to reduce the implementation complexity \cite{Oswald_2004, Ai_2011}, but their dependency on other errors has not been investigated in these studies. While these techniques have demonstrated efficacy for OFDM based systems, they can require substantial computational resources for estimation in systems that do not rely on the fast Fourier transform (FFT). Additionally, many SFO estimation methods assume perfect synchronization of the first samples and, therefore, do not account for sampling time offset (STO), which, in conjunction with SFO, introduces an additional challenge for accurate estimation. Apart from this, existing techniques typically work on a complex-valued signal, consequently increasing the number of real multiplications\footnote{\label{foo:complex} One complex multiplication requires a minimum of three real multiplications \cite{Fam_1988}.} required for the estimation.

Digital compensation of the SFO can be implemented in both the frequency and time domains \cite{Tsai_2005, Hamila_1998, Hu_2018, Liu_2024, Gao_2024}, where the latter is more efficient, especially for general bandlimited signals.  Existing methods typically carry out the SFO compensation by interpolating the signal using the Farrow structure \cite{farrow1998, Hamila_1998, Hu_2018, Gao_2024}. However, to the best of the authors' knowledge, the Farrow structure has not been explored for the SFO estimation before.

This paper introduces a time-domain SFO estimator that capitalizes on the Farrow structure that is used for the SFO compensation. The advantage of this approach is four-fold. Firstly, the implementation complexity of the estimator is reduced by leveraging the Farrow structure that is already used for the compensation. In addition, the proposed method utilizes only the real or imaginary part of a complex signal, which offers additional significant complexity reductions (also see Footnote \ref{foo:complex}).  Secondly, the proposed method works for arbitrary bandlimited signals, and thus imposes no additional constraints on the waveform structure, and there is no need for the FFT computation inherent in frequency-domain methods. Thirdly, our method handles the presence of STO and, importantly, enables joint SFO and STO estimation even in the presence of other synchronization errors, like CFO. Consequently, after SFO and STO estimation and compensation, any available CFO estimation and compensation method can be applied \cite{Singh_2025}.  
Lastly, in the proposed method for estimating two parameter (SFO and STO), which utilizes Newton's method, an efficient way of computing derivatives of the cost function via accumulators is introduced. This significantly reduces the implementation complexity by eliminating considerable parts of the multiplications.

Following this introduction, Section \ref{sec:compensation} gives a brief overview of the SFO problem and its compensation based on the Farrow structure. In Section \ref{sec:estimator}, the proposed time-domain estimation method is introduced, followed by its implementation in Section \ref{sec:complexity}.
Simulation results showing the application and performance of the proposed estimator are presented in Section \ref{sec:results}, whereas Section \ref{sec:conclusions} concludes the paper.
\section{SFO Compensation Using the Farrow Structure}\label{sec:compensation}
\vspace{\yspace}
\subsection{SFO Model}
Let $x_a(t)$ represent a bandlimited continuous-time signal to be sampled with slightly different sampling frequencies, $f_0$ and $f_1 = f_0 + \Delta_{f}$, where $\Delta_{f}$ denotes the SFO between them. The sampled signals can be expressed as
\begin{align}
	x_0(n)=x_a(nT), \quad x_1(n)=x_a(n(1+\Delta)T+\varepsilon), 
	\label{eq:problem_offset}
\end{align}
where $T = 1/f_0$ is the sampling period of the reference signal,  $\Delta=-\Delta_{f}/(f_0+\Delta_{f})$ represents the difference between the sampling periods, and $\varepsilon$ is the STO between the two signals. It is important to note that while the SFO, $\Delta_{f}$, and thereby $\Delta$, remain constant over time, the time deviation, $n\Delta T$ in \eqref{eq:problem_offset}, between samples $x_0(n)$ and $x_1(n)$ accumulates with increasing sample index \(n\).
The challenge here is to accurately estimate $\Delta$ and $\varepsilon$ to adequately compensate for these discrepancies.

\subsection{SFO Compensation Using the Farrow Structure}
\begin{figure}[tbp]
	\centering \includegraphics[scale=1.0]{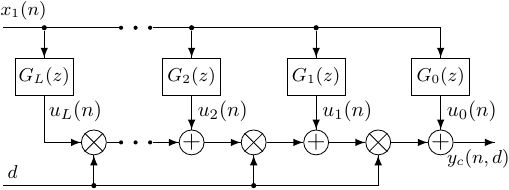}
	\caption{Fractional delay filter based on the Farrow structure.} 
	\label{Flo:farrow-equivalence-scheme}
\end{figure}

The Farrow structure \cite{farrow1998}, depicted in Fig. \ref{Flo:farrow-equivalence-scheme}, allows for adjusting a time-varying fractional delay $d$ in real time, ensuring high accuracy and efficient implementation due to the use of fixed-coefficient linear-phase FIR filters $G_k(z)$ \cite{Valimaki_1995, johansson2003}. In the context of the SFO, the fractional delay $d$ can be represented in terms of $\Delta$ and $\varepsilon$ as
\begin{equation}
	d(n, \Delta, \varepsilon)= n\Delta + \varepsilon,
	\label{eq:d}
\end{equation}
with the compensated signal given by
\begin{equation}
y_c(n, \Delta, \varepsilon) = \sum_{k=0}^{L} d^k(n, \Delta, \varepsilon) u_k(n),
\label{eq:farrow_output}
\end{equation}
where $u_k(n) = x_1(n) * g_k(n)$ represents the output of subfilter $G_k(z)$, with $*$ denoting convolution. Further, throughout the paper, $d$ is used instead of $d(n, \Delta, \varepsilon)$ to simplify the expressions.
\section{Proposed Time-Domain SFO Estimation Based on the Farrow Structure and Newton Method}
\vspace{\yspace}
\label{sec:estimator}
The proposed estimation method employs the output of the Farrow structure $y_c(n, \Delta, \varepsilon)$ in \eqref{eq:farrow_output}. Specifically, the objective function is expressed as the scaled-by-half\footnote{\label{foo:fraction}The factor $1/2$ is used in \eqref{eq:cost} for notation simplicity; it cancels out when taking the derivatives of the cost function.} squared error (SE) between the compensated signal $y_c(n, \Delta, \varepsilon)$ and the reference signal $x_0(n)$ as 
\begin{align}
    F(\Delta, \varepsilon) = \frac{1}{2} \sum_{n=0}^{N-1} \Big(y_c(n, \Delta, \varepsilon) - x_0(n) \Big)^2.
	\label{eq:cost}
\end{align}
The objective is to estimate the parameters ${\Delta}$ and ${\varepsilon}$ by minimizing $F(\Delta, \varepsilon)$, thereby solving the optimization problem given by
\begin{equation}
	\widehat{\Delta}, \widehat{\varepsilon} = \arg\min_{\Delta, \varepsilon} F(\Delta, \varepsilon).
\end{equation}
Given that $F(\Delta, \varepsilon)$, constructed as in \eqref{eq:cost}, with $y_c(n, \Delta, \varepsilon)$ as in \eqref{eq:farrow_output}, is twice differentiable, an efficient approach for finding its minimum through iterative updates is the well-known Newton's method \cite{boyd2004convex} (a.k.a. Newton-Raphson). This method, that offers rapid convergence when close to a local optimum \cite{kochenderfer2019algorithms}, is implemented through the update rule 
\begin{equation}
	\mathbf{w}^{(m+1)} = \mathbf{w}^{(m)} -\big(\mathbf{H}^{(m)}\big)^{-1} \mathbf{g}^{(m)},
	\label{eq:update}
\end{equation}
where $\mathbf{w} = [\Delta, \varepsilon]^{\top}$. Here $\mathbf{g}=\nabla F(\Delta, \varepsilon)$ stands for the gradient of the cost function defined in \eqref{eq:cost}, $\mathbf{H}=\nabla^2 F(\Delta, \varepsilon)$ for its Hessian matrix, and $m$ for the iteration index. To facilitate the computation of the updates, both the gradient and the Hessian matrix are expressed as functions of $d$ according to \big(recall that $\frac{\partial F}{\partial \Delta} =\frac{\partial F}{\partial d} \frac{\partial d}{\partial \Delta}$ and $\frac{\partial F}{\partial \varepsilon} =\frac{\partial F}{\partial d} \frac{\partial d}{\partial \varepsilon}$\big)
\begin{align}
	\hspace{-0.15cm}
   \mathbf{H}= \begin{bmatrix}
        \displaystyle \frac{\partial^2 F}{\partial \Delta^2} & \displaystyle \frac{\partial^2 F}{\partial \Delta \partial\varepsilon} \\ \\
        \displaystyle  \frac{\partial^2 F}{\partial\varepsilon \partial \Delta} & \displaystyle\frac{\partial^2 F}{\partial \varepsilon^2}
    \end{bmatrix}
    & =\begin{bmatrix}
    \displaystyle\sum_{n} n^2 \frac{\partial^2 F_n}{\partial d^2} & \displaystyle\sum_{n} n \frac{\partial^2 F_n}{\partial d^2} \\ \\
    \displaystyle\sum_{n}   n \frac{\partial^2 F_n}{\partial d^2} & \displaystyle\sum_{n}   \frac{\partial^2 F_n}{\partial d^2}
    \end{bmatrix}\hspace{-0.12cm},\label{eq:Hessian} \\
	\mathbf{g} =  
    \begin{bmatrix} 
        \displaystyle\frac{\partial F}{\partial \Delta} &
        \displaystyle\frac{\partial F}{\partial \varepsilon}
    \end{bmatrix}^{\top}
	&=\begin{bmatrix} 
		\displaystyle\sum_{n}n\frac{\partial F_n}{\partial d} &
		\displaystyle\sum_{n}\frac{\partial F_n}{\partial d} 
	\end{bmatrix}^{\top}\hspace{-0.1cm}, \label{eq:gradient}
\end{align}
where  $F=F(\Delta, \varepsilon)=\sum_n F_n$ as in \eqref{eq:cost}, with $F_n=1/2\big(y_c(n, \Delta, \varepsilon) - x_0(n)\big)^2$ being  used for notation simplicity. Here, the first- and second-order derivatives of $F_n$ are
\begin{align}
	\frac{\partial F_n}{\partial d} &=  \Bigg(\sum_{k=0}^{L} d^k u_k(n)-x_0(n)\Bigg) \Bigg(\sum_{k=1}^{L} k d^{k-1} u_k(n)\Bigg), \hspace{-3.0 pt} \\
	\frac{\partial^2 F_n}{\partial d^2} &= \Bigg(\sum_{k=1}^{L} k d^{k-1} u_k(n)\Bigg)^2 + \Bigg(\sum_{k=0}^{L} d^k u_k(n)-x_0(n)\Bigg) \nonumber \\ 
	&\times\Bigg(\sum_{k=2}^{L} k(k-1) d^{k-2} u_k(n)\Bigg).
\end{align}
\section{Low-Complexity Implementation} \label{sec:complexity}
\begin{figure*}[tbp]
    \centering
	\centering \includegraphics[scale=1.0]{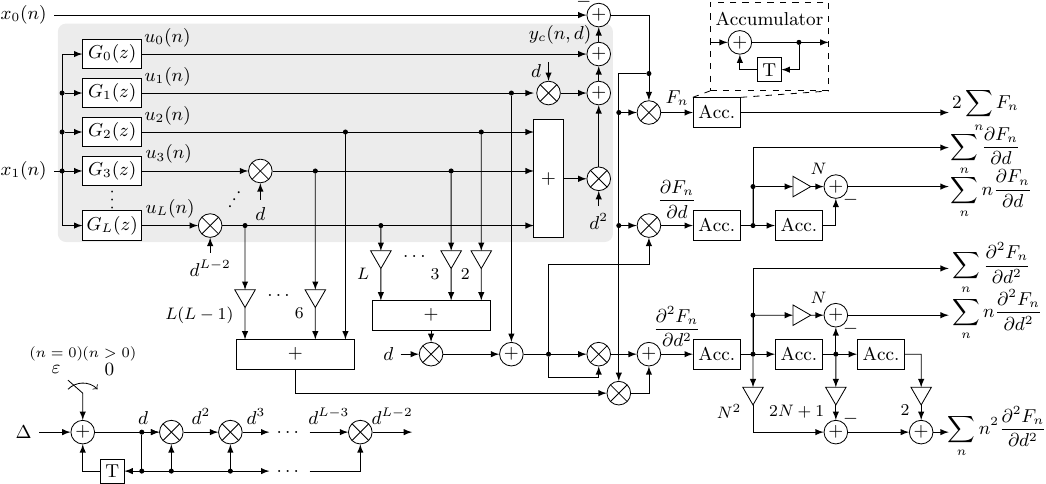}
    \caption{Implementation of the SFO compensation (highlighted with the gray rectangular) and derivatives required for estimation.}\label{Flo:derivatices}
\end{figure*}

\vspace{\yspace}
The estimator can be implemented cost-efficiently with the computations of the derivatives presented in Fig. \ref{Flo:derivatices}. Furthermore, we introduce an efficient method below for calculating the elements in \eqref{eq:Hessian} and \eqref{eq:gradient}.
\vspace{-3 pt}
\subsection{Time-Index and Time-Index-Squared Weighted Sum Computations Based on Cascaded Accumulators} \label{sec:accumulators}
To efficiently calculate the time-indexed weighted (TIW) sums, $\sum_n n \frac{\partial F_n}{\partial d}$ and $\sum_n n \frac{\partial^2 F_n}{\partial d^2}$, and time-index-squared weighted (TISW) sum, $\sum_n n^2 \frac{\partial^2 F_n}{\partial d^2}$, in \eqref{eq:Hessian}--\eqref{eq:gradient}, related to the first and second derivatives, we propose an efficient implementation based on cascaded accumulators. Denoting the input to the first accumulator $v_n= \frac{\partial^2 F_n}{\partial d^2}$, the second derivative in \eqref{eq:Hessian}, it can be shown\footnote{Due to a lack of space, the proof of \eqref{eq:accum2}--\eqref{eq:S3} is omitted in this paper.} that the output of the first, second, and third cascaded accumulators can be expressed in terms of the input $v_n$ as
\begin{align}
	A_1(N) &= \sum_{n=0}^{N-1} v_n, \quad A_2(N) = \sum_{n=0}^{N-1} (N-n)v_n, \quad \label{eq:accum2}\\
	A_3(N) &= \sum_{n=0}^{N-1} \frac{(N-n)(N-n+1)}{2} v_n.
\end{align}
To generate the matrix elements in \eqref{eq:Hessian}, the output of the first accumulator, $A_1(N)$, directly gives the sum of $v_n$, while the TIW and TISW sums are computed by utilizing the outputs of the accumulators scaled by constants as
\begin{align}
	\sum_{n=0}^{N-1} n v_n   &= NA_1(N) - A_2(N),  \label{eq:S2} \\ 
	\sum_{n=0}^{N-1} n^2 v_n & = N^2 A_1(N) - (2N+1)A_2(N) +2A_3(N). \label{eq:S3}
\end{align} 
Likewise, the elements in \eqref{eq:gradient} can be obtained using two cascaded accumulators, with the input to the first accumulator set as $v_n = \frac{\partial F_n}{\partial d}$ and with the outputs computed as in \eqref{eq:accum2}. Then the desired TIW sum can be calculated as in \eqref{eq:S2}.
\subsection{Implementation Complexity of the Proposed Estimator}
The implementation complexity of the proposed estimator comprises the update computations as described in \eqref{eq:update}, with the subsequent matrix inversion, and the computation of the derivatives to obtain the elements in \eqref{eq:Hessian}--\eqref{eq:gradient}, via the proposed method for the sum of multiplications using cascaded accumulators outlined in Section \ref{sec:accumulators}. To assess the complexity, we distinguish between general multiplications and constant multiplications since the latter can be efficiently implemented using adders and shift operators. Additionally, some parts of the estimation algorithm, specifically the subfilter outputs $u_k(n)$, $k = 0, \dots, L$, multiplications by $d$, and additions required to compute the output $y_c(n,d)$, are utilized for compensation (in Fig.~\ref{Flo:derivatices}, this part is highlighted with a gray rectangle). Given that they are inherently required for this purpose, their cost is excluded from the overall complexity of the estimator. Therefore, considering that every accumulator in \eqref{eq:S2}--\eqref{eq:S3} needs $N-1$ additions, the proposed estimation algorithm requires $(L+2)\times N + 8$ general multiplications, $(2L-3)\times N+5$ real multiplications, $(2L+6)\times N+3$ additions, and one division per batch of $N$ samples. This includes the cost of the update in \eqref{eq:update}, which totally requires eight general multiplications, five additions and one division. Further in regular Farrow structures, $L$ is a small number, typically between one and five.

It is important to highlight that the proposed estimator, which relies on the Newton's method, requires only a few iterations to converge, and thus the total implementation complexity per sample is low. Moreover, unlike frequency-domain estimation methods, the proposed algorithm does not require the FFT computation, thus the number of arithmetic operations is reduced.

\subsection{A Note on the Implementation of the Compensator}
While the proposed estimation algorithm operates only on one of the components in case of a complex-valued signal, the compensation must be applied to both components. Due to the fact that processing of the real and imaginary parts are implemented as two separate branches in practice (since the Farrow filter impulse responses are real-valued), two instances of the Farrow structure are required for the SFO correction. It is also noted again that such compensation is needed in all SFO correction techniques independently of the estimation method.
\section{Results} \label{sec:results}
In this section, we evaluate the performance of the proposed method by applying it to different types of bandlimited signals. The goal is to synchronize two signals sampled at $f_1$ and $f_0$, where $f_0$ serves as a reference with a normalized sampling frequency of one. Examples 1--3 demonstrate specific instances from evaluation across $1\,000$ signals for each case. For the SFO and STO set as $\Delta=-200$ ppm and $\varepsilon = 0.03$, respectively, the estimation error is upper bounded by a $3$\% error margin around the target values, with the majority ($\approx
90$\%) of the cases achieving estimation errors upper bounded by $1$\%. The signal-to-noise ratio (SNR) is $60$ dB, and $N=256$ samples are used for the estimation.

\textit{Example 1:} A multi-sine real-valued signal, generated with phases and amplitudes from the 16-QAM scheme, is considered here. Applying the proposed method, it suffices to have one iteration, and the values are found to be \mbox{$\widehat{\Delta}=-199.73$ ppm} and $\widehat{\varepsilon}=0.0301$, which are close to the actual values. 

\begin{figure}[tbp!]
	\centering
	\centering \includegraphics[scale=1.0]{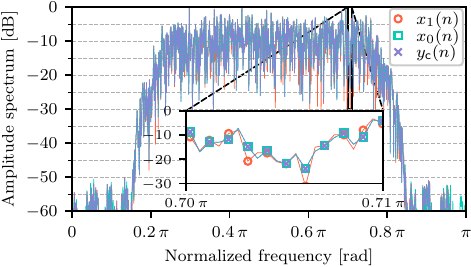}
	\caption{\textit{Example 2: }Amplitude spectrum before and after compensation using the Farrow structure for a bandpass-filtered white noise signal.}
	\label{Flo:noise}
\end{figure}

\textit{Example 2:} Here, a bandpass-filtered white noise signal is considered, the estimator converges also after one iteration with \mbox{$\widehat{\Delta}=-199.97$} ppm and $\widehat{\varepsilon}=0.0299$.
In Fig.~\ref{Flo:noise}, the reference signal $x_0(n)$, SFO-affected signal $x_1(n)$, and compensated signal $y_c(n)$ are shown, where the latter aligns with the reference signal, that together with the estimated parameters demonstrate accuracy for this type of signal.

\vspace{-3px}
\begin{figure}[tbp!]
	\centering
	\centering \includegraphics[scale=1.0]{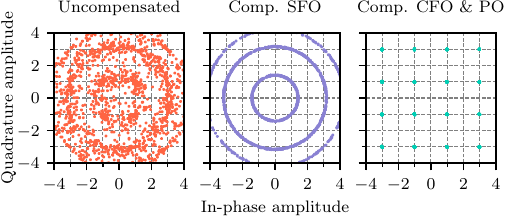}
	\caption{\textit{Example 3:} Constellation diagram: (a) uncompensated, (b) after SFO estimation and compensation, and (c) after CFO and PO compensation.}
	\label{Flo:constellation}
\end{figure}
\vspace{-3px}

\textit{Example 3:} To illustrate the effectiveness of the proposed method in the presence of other errors, $5\%$ CFO and phase offset (PO) are introduced to an OFDM signal with $1\,536$ out of $2\,048$ subcarriers, and 16-QAM. Here, the estimation of SFO and STO also requires one iteration to obtain \mbox{$\widehat{\Delta}=-201.03$ ppm} and $\widehat{\varepsilon}=0.0298$, by using only the real component of the signal for the SFO estimation. Figure~\ref{Flo:constellation} shows the constellation diagrams before, after SFO compensation, and after subsequent CFO and PO compensation. 

\begin{figure}[tbp!]
	\centering
	\centering \includegraphics[scale=1.0]{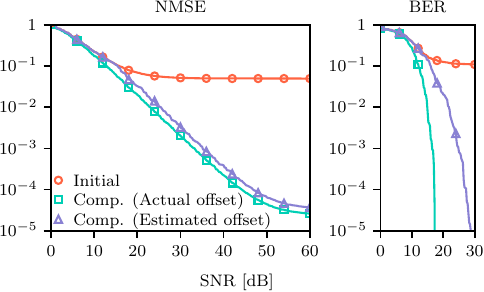}
	\caption{\textit{Example 4: }NMSE versus SNR using the proposed estimator for OFDM signals.}
	\label{Flo:MSE}
\end{figure}

\textit{Example 4:} Further, Fig.~\ref{Flo:MSE} quantifies the performance of the estimator by plotting the normalized mean-squared error (NMSE) and bit error rate (BER) achieved versus SNR for $1\,000$ OFDM signals with ${\Delta}=-200$ ppm and ${\varepsilon}=300$ ppm per SNR value, demonstrating the estimator’s effectiveness even in presence of low SNRs. As in previous examples, one iteration is sufficient.
\vspace{-3px}

\begin{figure}[tbp!]
	\centering
	\centering \includegraphics[scale=1.0]{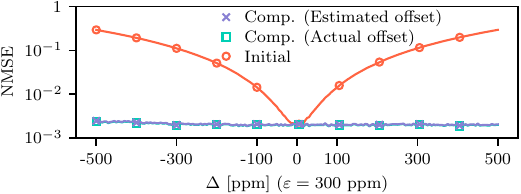}
	\caption{\textit{Example 5: }NMSE versus $\Delta$ using the proposed estimator for OFDM signals.}
	\label{Flo:MSE_delta}
\end{figure}

\textit{Example 5:} Lastly, the performance of the proposed estimator within a wide range of SFOs is analyzed. Here, OFDM signals using 16-QAM are generated with fixed $\varepsilon=300$ ppm and varying $\Delta$. Figure~\ref{Flo:MSE_delta} shows the NMSE versus $\Delta$ across $1\,000$ signals per $\Delta$ value, demonstrating the estimator’s performance under different SFOs. It is seen that the compensated signals are close to the reference. Here, the NMSE is bounded by an SNR of $30$ dB.
\vspace{-3px}
\section{Conclusions\label{sec:conclusions}}
\vspace{\yspace}
This paper proposed a low-complexity time-domain SFO estimator based on the Farrow structure, which is also used for SFO compensation. 
The estimation algorithm uses the Newton's method that ensures rapid convergence and accurate estimation. 
Through the simulation examples, the proposed method demonstrated its capability of estimating SFO for any type of bandlimited signals, and its robustness against other impairments, highlighting its potential for practical applications. A detailed comparison with frequency domain estimators will be explored in future works.
\vspace{-2px}
\section*{Acknowledgment}
\vspace{-2px}
This work is financially supported by ELLIIT and Sweden's Innovation Agency.

\newpage
\begin{small} 
	\bibliographystyle{IEEEtran}
   \bibliography{IEEEabrv, references/bibliography} 

\begin{thebibliography}{10}
\providecommand{\url}[1]{#1}
\csname url@samestyle\endcsname
\providecommand{\newblock}{\relax}
\providecommand{\bibinfo}[2]{#2}
\providecommand{\BIBentrySTDinterwordspacing}{\spaceskip=0pt\relax}
\providecommand{\BIBentryALTinterwordstretchfactor}{4}
\providecommand{\BIBentryALTinterwordspacing}{\spaceskip=\fontdimen2\font plus
\BIBentryALTinterwordstretchfactor\fontdimen3\font minus
  \fontdimen4\font\relax}
\providecommand{\BIBforeignlanguage}[2]{{%
\expandafter\ifx\csname l@#1\endcsname\relax
\typeout{** WARNING: IEEEtran.bst: No hyphenation pattern has been}%
\typeout{** loaded for the language `#1'. Using the pattern for}%
\typeout{** the default language instead.}%
\else
\language=\csname l@#1\endcsname
\fi
#2}}
\providecommand{\BIBdecl}{\relax}
\BIBdecl

\bibitem{Matin_2021}
S.~A. Matin and L.~B. Milstein, ``{OFDM} system performance, variability and
  optimality with design imperfections and channel impediments,'' \emph{IEEE
  Trans. Veh. Tech.}, vol.~70, no.~1, pp. 381--397, 2021.

\bibitem{Wang_2003}
X.~Wang, T.~Tjhung, Y.~Wu, and B.~Caron, ``{SER} performance evaluation and
  optimization of {OFDM} system with residual frequency and timing offsets from
  imperfect synchronization,'' \emph{IEEE Trans. Broadcast.}, vol.~49, no.~2,
  pp. 170--177, 2003.

\bibitem{Hou_2020}
Y.~Hou, X.~Chen, and K.~Du, ``Comparison and analysis of {SFO} estimation
  methods based on {OFDM},'' in \emph{5th Int. Conf. Comput. Commun. Syst.
  (ICCCS)}, 2020, pp. 640--644.

\bibitem{Li_2024}
L.~Li, Z.~Zhan, Z.~Pang, Q.~Yang, X.~Liu, B.~Huang, and S.~Wei, ``A sampling
  frequency offset estimation method for high-speed power line communication
  system,'' \emph{9th Int. Conf. Comp. Communic. Syst. (ICCCS)}, pp.
  1094--1099, 2024.

\bibitem{Nguyen_2009}
H.~Nguyen-Le, T.~Le-Ngoc, and C.~C. Ko, ``{RLS}-based joint estimation and
  tracking of channel response, sampling, and carrier frequency offsets for
  {OFDM},'' \emph{IEEE Trans. Broadcast.}, vol.~55, no.~1, pp. 84--94, 2009.

\bibitem{Tsai_2005}
P.-Y. Tsai, H.-Y. Kang, and T.-D. Chiueh, ``Joint weighted least-squares
  estimation of carrier-frequency offset and timing offset for {OFDM} systems
  over multipath fading channels,'' \emph{IEEE Trans. Veh. Technol.}, vol.~54,
  no.~1, pp. 211--223, 2005.

\bibitem{Kim_2011}
Y.-H. Kim and J.-H. Lee, ``Joint maximum likelihood estimation of carrier and
  sampling frequency offsets for {OFDM} systems,'' \emph{IEEE Trans.
  Broadcast.}, vol.~57, no.~2, pp. 277--283, 2011.

\bibitem{Xu2014}
X.~Wang and B.~Hu, ``A low-complexity {ML} estimator for carrier and sampling
  frequency offsets in {OFDM} systems,'' \emph{IEEE Commun. Lett.}, vol.~18,
  no.~3, pp. 503--506, 2014.

\bibitem{WU2024}
Y.~Wu, R.~Mei, and J.~Xu, ``Non pilot data-aided carrier and sampling frequency
  offsets estimation in fast time-varying channel,'' \emph{Big Data Res.},
  vol.~36, p. 100461, 2024.

\bibitem{Oswald_2004}
E.~Oswald, ``{NDA} based feedforward sampling frequency synchronization for
  {OFDM} systems,'' \emph{Proc., IEEE 59th. Veh. Technol. Conf. (VTC-Spring)},
  vol.~2, pp. 1068--1072, 2004.

\bibitem{Ai_2011}
B.~Ai, Y.~Shen, Z.~D. Zhong, and B.~H. Zhang, ``Enhanced sampling clock offset
  correction based on time domain estimation scheme,'' \emph{IEEE Trans.
  Consum. Electron.}, vol.~57, no.~2, pp. 696--704, 2011.

\bibitem{Fam_1988}
A.~Fam, ``Efficient complex matrix multiplication,'' \emph{IEEE Trans.
  Comput.}, vol.~37, no.~7, pp. 877--879, 1988.

\bibitem{Hamila_1998}
R.~Hamila, J.~Vesma, H.~Vuolle, and M.~Renfors, ``Effect of frequency offset on
  carrier phase and symbol timing recovery in digital receivers,'' \emph{Proc.
  {URSI} Int. Symp. Signals Syst., and Electron. Conf.}, pp. 247--252, 1998.

\bibitem{Hu_2018}
Q.~Hu, X.~Jin, and Z.~Xu, ``Compensation of sampling frequency offset with
  digital interpolation for {OFDM}-based visible light communication systems,''
  \emph{J. Light. Technol.}, vol.~36, no.~23, pp. 5488--5497, 2018.

\bibitem{Liu_2024}
Y.~Liu, W.~Yang, Z.~Wang, and Y.~Xu, ``Easy-hardware-implementation algorithm
  of carrier and sampling frequency offset estimation in {OFDM} systems,'' in
  \emph{OES China Ocean Acoust. (COA)}, 2024, pp. 1--4.

\bibitem{Gao_2024}
X.~Gao, M.~Chen, J.~Zhou, J.~Xu, H.~Li, Y.~Cai, and D.~Wang, ``Sampling
  frequency offset estimation and compensation for asynchronous optical {IMDD}
  {FBMC} systems,'' \emph{J. Light. Technol.}, vol.~42, no.~9, pp. 3099--3106,
  2024.

\bibitem{farrow1998}
C.~Farrow, ``A continuously variable digital delay element,'' \emph{Proc. IEEE
  Int. Symp. Circuits Syst.}, vol.~3, pp. 2641--2645, 1988.

\bibitem{Singh_2025}
S.~Singh, S.~Kumar, S.~Majhi, U.~Satija, and C.~Yuen, ``Blind carrier frequency
  offset estimation techniques for next-generation multicarrier communication
  systems: Challenges, comparative analysis, and future prospects,'' \emph{IEEE
  Commun. Surv. Tutor}, vol.~27, no.~1, pp. 1--36, 2025.

\bibitem{Valimaki_1995}
V.~V\"alimaki, ``A new filter implementation strategy for {L}agrange
  interpolation,'' \emph{Proc. IEEE Int. Symp. Circuits Syst.}, pp. 361--364,
  1995.

\bibitem{johansson2003}
H.~Johansson and P.~L\"owenborg, ``On the design of adjustable fractional delay
  {FIR} filters,'' \emph{IEEE Trans. Circuits Syst. II: Express Briefs},
  vol.~50, no.~4, pp. 164--169, 2003.

\bibitem{boyd2004convex}
S.~Boyd and L.~Vandenberghe, \emph{Convex optimization}.\hskip 1em plus 0.5em
  minus 0.4em\relax Cambridge university press, 2004.

\bibitem{kochenderfer2019algorithms}
M.~Kochenderfer and T.~Wheeler, \emph{Algorithms for Optimization}.\hskip 1em
  plus 0.5em minus 0.4em\relax MIT Press, 2019.

\end{thebibliography}
\end{small}

\end{document}